\documentclass[final,3p,times,twocolumn]{elsarticle}
\usepackage{graphicx}
\usepackage{amssymb}
\usepackage{amsthm}
\newcommand{\be}{\begin{equation}}
\newcommand{\ee}{\end{equation}}
\journal{Physics Letters A}
 
\begin{document}
\begin{frontmatter}

\title{Fractional Standard Map}

\author[ci,yu]{Mark Edelman\corref{cor}\fnref{me}}
\ead{edelman@cims.nyu.edu}
\author[ci,si]{Vasily E. Tarasov}
%\author[ci]{Harold Weitzner}
\cortext[cor]{Corresponding author}
\address[ci]{Courant Institute of Mathematical Sciences,
New York University, 251 Mercer St., New York, NY 10012, USA}
\address[yu]{Department of Physics, Stern College at Yeshiva University,
245 Lexington Ave, New York, NY 10016, USA} 
\address[si]{Skobeltsyn Institute of Nuclear Physics,
Moscow State University, Moscow 119991, Russia}

%\date{\today}

\begin{abstract}
Properties of the phase space of the standard map with memory are investigated.
This map was obtained from a kicked fractional differential equation. 
Depending on the value of the map parameter and the fractional
order of the derivative in the original differential equation, this
nonlinear dynamical system demonstrates attractors (fixed points, 
stable periodic trajectories, slow converging and slow diverging 
trajectories, ballistic trajectories, and fractal-like structures) and/or 
chaotic trajectories. At least one type of fractal-like sticky attractors 
%in the chaotic sea 
was observed.  
 \end{abstract}

\end{frontmatter}

\section{Introduction}

The standard map (SM) can be derived from the differential equation 
describing kicked rotator. The description of many physical systems and 
effects (Fermi acceleration, comet dynamics, etc.) can be reduced   
to the studying of the SM \cite{Chirikov}. 
The SM provides the simplest model of the
universal generic area preserving map and it is one of the 
most widely studied maps. The topics examined include 
fixed points, elementary structures of islands and a chaotic sea,
and fractional kinetics \cite{Chirikov,LichLib,Zaslavsky1}. 

It was recently realized that many physical systems, including
systems of oscillators with long range interaction
\cite{TZ3,JPA2006}, non-Markovian systems 
with memory (\cite{Podlubny} Ch.10, \cite{Mem1,Mem2,Mem3,Mem4,Mem5}), 
fractal media \cite{Mainardi}, etc., can be described by the fractional 
differential equations (FDE)  \cite{Podlubny,SKM,KST}.
As with the usual differential equations, the reduction of FDEs
to the corresponding maps can provide a valuable tool for the 
analysis of the properties of the original systems.
As in the case of the SM, the fractional standard map (FSM), 
derived in \cite{Main} from the fractional differential equation 
describing a kicked system,
is perhaps the best candidate
to start a general investigation of the properties of 
maps which can be obtained from FDEs. 

As it was shown in  \cite{Main}, maps that can be derived from FDEs
are of the type of discrete maps with memory. One-dimensional maps with 
memory,  in which
the present state of evolution depends on all past states,  
studied previously  \cite{Ful,Fick1,Fick2,Giona,Gallas,Stan}
were not derived from differential equations. 
Most results were obtained 
for the generalizations of the logistic map. 

In the physical systems the transition from integer order time
derivatives to fractional (of a lesser order) introduces additional damping 
and is similar in appearance to additional friction \cite{Podlubny,ZSE}. 
Accordingly, in the case of the FSM we may
expect transformation of the islands of stability and the accelerator mode
islands 
into attractors (points, attracting trajectories, strange attractors).
Because the damping in systems with fractional derivatives is based on the
internal causes different from the external forces of friction 
\cite{ZSE,Stan2},
the corresponding attractors are also different from the attractors of the 
regular systems with friction and are called fractional attractors \cite{ZSE}. 
Even in one-dimensional cases \cite{Ful,Fick1,Fick2,Giona,Gallas,Stan} 
most of the results were obtained numerically. An additional dimension 
makes the problem even more complex and most of the results in the present
paper were obtained numerically.

\section{FSM, initial conditions}

The standard map in the form

\be 
%\begin{gathered}
p_{n+1} = p_n - K \sin x_n, \ \  x_{n+1} = x_n + p_{n+1}  \  \ ({\rm mod} \ 2\pi )
\label{1}
\ee 
%$$
%\be
%x_{n+1} = x_n + p_{n+1}  \ \ \ \ ({\rm mod} \ 2\pi )
%\end{gathered}
%\label{1}
%\ee
can be derived from the differential equation
\be
\ddot{x}+K \sin(x) \sum^{\infty}_{n=0} \delta \Bigl(\frac{t}{T}-n \Bigr)=0.
\label{1d}
\ee

By replacing the second-order time derivative in eq. (\ref{1d})
with the Riemann-Liouville derivative $ _0D^{\alpha}_t$ 
one obtains a fractional equation of the motion in the form
%Replacing in (\ref{1d}) fractional Riemann-Liouville derivative,
%we obtain
%$ _0D^{\alpha}_t$ for the second order derivative
\be \label{1f}
_0D^{\alpha}_t x+K\sin(x) \sum^{\infty}_{n=0} \delta \Bigl(\frac{t}{T}-n \Bigr)=0, 
\quad (1 <\alpha \le 2) ,
\ee
where
$$
_0D^{\alpha}_t x(t)=D^n_t \ _0I^{n-\alpha}_t x(t)=
$$
\be
\frac{1}{\Gamma(n-\alpha)} \frac{d^n}{dt^n} \int^{t}_0 
\frac{x(\tau) d \tau}{(t-\tau)^{\alpha-n+1}}  \quad (n-1 <\alpha \le n),
\label{RLFD}
\ee
$D^n_t=d^n/dt^n$, and $ _0I^{\alpha}_t$ is a fractional integral.
The initial conditions for  (\ref{1f}) are
$$
%\begin{gathered}
(_0D^{\alpha-1}_tx) (0+) = p_1, \\
$$
\be
(_0D^{\alpha-2}_tx) (0+) = b.
%\end{gathered}
\label{1ic}
\ee
The Cauchy type problem (\ref{1f}) and (\ref{1ic}) is equivalent to the
Volterra  integral equation of the second kind \cite{KBT1,KBT2,TarV}
$$
x(t)= \frac{p_1}{\Gamma(\alpha)} t^{\alpha-1} +
\frac{b}{\Gamma(\alpha - 1)} t^{\alpha-2} 
$$
\be
- \frac{K}{\Gamma(\alpha)} 
\int^t_0 \frac{\sin[x(\tau)] \sum^{\infty}_{n=0} \delta
  \Bigl(\frac{\tau}{T}-n \Bigr) d\tau}{(t-\tau)^{1-\alpha}}.
\label{RL6b}
\ee
Defining the momentum as
\be \label{1m}
p(t)= \, _0D^{\alpha-1}_t x(t),
\ee
and performing integration in (\ref{RL6b})
one can derive the equation for the FSM in the form
(for the thorough derivation see  \cite{TarV})
\be \label{2}
p_{n+1} = p_n - K \sin x_n ,
%\eqno (2)
\ee
%\be
$$
x_{n+1} = \frac{1}{\Gamma (\alpha )} 
\sum_{i=0}^{n} p_{i+1}V_{\alpha}(n-i+1) +
$$
\be \label{3}
\frac{b}{\Gamma(\alpha-1)} (n+1)^{\alpha-2}
, \ \ \ \ ({\rm mod} \ 2\pi ) ,
%\eqno (3)
\ee
where 
\be \label{4}
V_{\alpha}(m)=m^{\alpha -1}-(m-1)^{\alpha -1}. 
%\eqno (4)
\ee
Here it is assumed that $T=1$ and $1<\alpha\le2$.
The form of eq.~(\ref{3}) which provides a more clear correspondence with
the SM ($\alpha=2$) in the case $b=0$ is presented 
in  Sec.~\ref{p1} (eq.~(\ref{5})).

The second initial condition in  (\ref{1ic}) can be written as
%\be
$$
_0D^{\alpha-2}_t (0+) = \lim_{t \rightarrow 0+}\, _0I^{2-\alpha}_t x(t)=
$$
\be \label{4ic}
 \lim_{t \rightarrow 0+}\frac{1}{\Gamma(2-\alpha)} \int^{t}_0 
\frac{x(\tau) d \tau}{(t-\tau)^{\alpha-1}}=b , \quad (1 <\alpha \le 2) ,
\ee
which requires $b=0$ in order to have a solution bounded at $t=0$ for
$\alpha<2$. The assumption $b=0$ leads to the FSM 
equations which in the limiting
case $\alpha=2$ coincide with the equations for the standard map under the
condition  $x_0=0$.

In this paper the FSM is taken in the form derived in \cite{Main} 
which coincides
with  (\ref{2}) and (\ref{3}) if $b=0$. 
It is also assumed that $x_0=0$ and the results can be compared
to those obtained for the SM with $x_0=0$ and arbitrary $p_0$.    
As a test,
for the SM and for the FSM with $\alpha=2$ and the same initial conditions 
numerical calculations show that phase portraits 
look identical.

System of equations (\ref{2}) and (\ref{3}) can be considered  
either in a cylindrical 
phase space ($x$  mod  $2 \pi$) or in  
unbounded phase space. The second case is convenient to study transport. 
The trajectories in the second case are easily related to the first case. 
The FSM has no periodicity in $p$ (the SM does) 
and cannot be considered on a torus.

\section{Stable fixed point}

The SM has stable fixed points at (0,$2\pi n$) for $ K<K_c=4$.
It is easy to see that point $(0,0)$ is also a fixed point for the FSM.
Direct computations using (\ref{2}) and (\ref{3})
demonstrate that for the small initial values
of $p_0$ there is a clear transition from the convergence to the fixed
point to divergence when the value of the parameter $K$ 
crosses the curve $K=K_c(\alpha )$ on Fig.~1a from smaller to larger values.

The following system describes the evolution of trajectories near fixed
point  $(0,0)$

\be \label{5n}
\delta p_{n+1} = \delta p_n - K \delta x_n ,
%\eqno (5)
\ee
\be \label{6n}
\delta x_{n+1} = \frac{1}{\Gamma (\alpha )} 
\sum_{i=0}^{n} \delta p_{i+1}V_{\alpha}(n-i+1) .
%\eqno (6)
\ee

\begin{figure}
\centering
\rotatebox{0}{\includegraphics[width=7 cm]{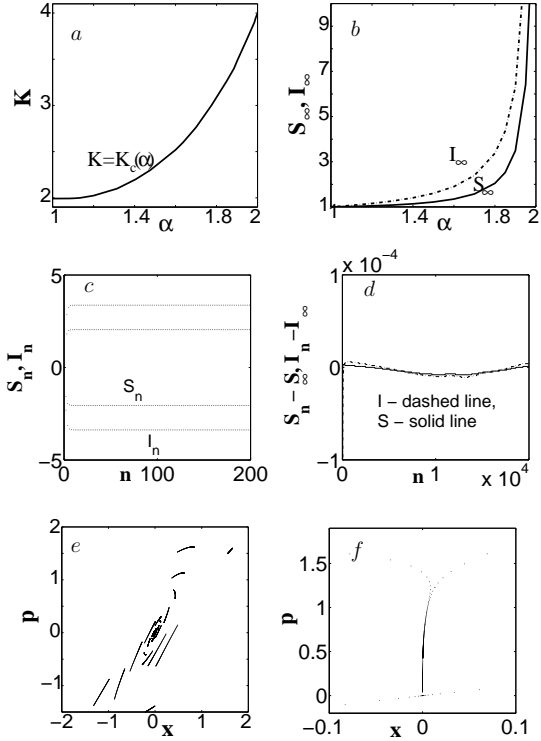}}
\caption{\label{Fig1}  Stability of the fixed point(0,0): a). The fixed
  point is stable below the curve $K=K_c(\alpha )$; b). Values of $S_\infty$ 
and $I_\infty$ obtained after 20000 iterations of eqs.  (\ref{15n}) and 
(\ref{16n}). As $\alpha \rightarrow 2$ the values $S_\infty$ 
and $I_\infty$ increase rapidly. For  $\alpha=1.999$,  $S_\infty \approx
276$ and $I_\infty \approx 552$ after 20000 iterations; 
c). An example of the typical evolution of $S_\infty$ and 
$I_\infty$ over the first 200 iterations for $1 < \alpha <2$.  
This particular figure corresponds to $\alpha =1.8$;
d). Deviation of the values $S_n$ and $I_n$ from the values 
$S_\infty \approx 2.04337$ and  $I_\infty \approx 3.37416$ for 
$\alpha =1.8$ during the first 20000 iterations (this type of behavior 
remains for  $1 < \alpha <2$); 
e). Evolution of
trajectories with $p_0=1.5+0.0005i$, $0 \le i < 200$ for the case $K=3$,
$\alpha=1.9$. The line segments correspond to the $n$th iteration on the
set of trajectories with close initial conditions. 
The evolution of the trajectories with smaller $p_0$ is 
similar; f). $10^5$ iterations on both of two trajectories for 
$K=2$, $\alpha =1.4$. The one at the bottom with $p_0=0.3$ is a fast
converging trajectory. The upper trajectory with  $p_0=5.3$ is an example
of the ASCT in which $p_{100000} \approx 0.042$.  
 }
\end{figure}
The solution 
can be found in the form
\be \label{7n}
\delta p_{n} = p_0\sum_{i=0}^{n-1}p_{n,i}\Bigl(\frac{2}{V_{\alpha
l}}\Bigr)^i\Bigl(\frac{V_{\alpha l}K}
{2 \Gamma (\alpha )}\Bigr)^i, \quad ( n > 0) ,
%\eqno (7)
\ee
\be \label{8n}
\delta x_{n} = \frac{p_0}{\Gamma (\alpha )}\sum_{i=0}^{n-1}x_{n,i}
\Bigl(\frac{2}{V_{\alpha l}}\Bigr)^i\Bigl(\frac{V_{\alpha l}K}{2 \Gamma
  (\alpha )}\Bigr)^i, \quad (n > 0),
%\eqno (8)
\ee
The origin of the terms in parentheses, as well as the definition 
\be \label{9n} 
 V_{\alpha l}  =  \sum_{k=1}^{\infty} (-1)^{k+1} V_{\alpha}(k)
\ee
will become clear in Sec.~\ref{p2}.
Eqs. (\ref{5n}) - (\ref{9n}) lead to the following iterative relationships
\be \label{10n}
x_{n+1,i}=-\sum_{m=i}^{n}(n-m+1)^{\alpha-1}x_{m,i-1} , \quad ( 0 <i \le n) ,
\ee
\be \label{11n}
p_{n+1,i}=-\sum_{m=i}^{n}x_{m,i-1} , \quad ( 0 <i < n)  
\ee
with the initial and boundary conditions
\be \label{12n}
%p_{n+1,n}=x_{n+1,n}=(-1)^n, \ \ p_{n+1,0}=1,  \quad x_{n+1,0}=(n+1)^{\alpha-1}.
p_{n+1,n}=x_{n+1,n}=(-1)^n, \quad p_{n+1,0}=1,  
\ee
$$
 x_{n+1,0}=(n+1)^{\alpha-1}.
$$
From  (\ref{10n}) and  (\ref{11n}) it is clear that the series 
(\ref{7n}) and  (\ref{8n}) are alternating and it is natural to apply the
Dirichlet's test to verify their convergence. This can be done by
considering the totals
\be \label{13n}
S_n=\sum_{i=0}^{n-1}x_{n,i}\Bigl(\frac{2}{V_{\alpha l}}\Bigr)^i,
\ee
\be \label{14n}
I_n=\sum_{i=0}^{n-1}p_{n,i}\Bigl(\frac{2}{V_{\alpha l}}\Bigr)^i.
\ee
They obey the following iterative rules
\be \label{15n}
S_n= n^{\alpha -1}-\frac{2}{V_{\alpha l}} \sum_{i=1}^{n-1}(n-i)^{\alpha -1}S_i,
\quad S_1=1,
\ee
\be \label{16n}
I_n=1 -\frac{2}{V_{\alpha l}} \sum_{i=1}^{n-1}S_i.
\ee
Computer simulations show that values of $S_n$ and $I_n$ converge 
to the values $(-1)^{n+1}S_\infty $ and $(-1)^{n+1}I_\infty $ depicted
on Fig.~1b. Figs.~1c,~1d show an example of the typical evolution of  
$S_n$ and $I_n$ 
%%% for $\alpha=1.8$ 
over the first 20000 iterations. 
%% There is still no mathematical proof of the convergence.
It means that the condition of convergence of $\delta p_{n}$ 
and  $\delta x_{n}$ is 
\be \label{17n}
\frac{V_{\alpha l}K}{2 \Gamma (\alpha )} <1.
\ee
Numerical evaluation of the equality $K= 2 \Gamma (\alpha )/V_{\alpha l}$
perfectly reproduces the curve on Fig.~1a obtained by the direct
 computations of (\ref{2}) and (\ref{3}).

Because not only the stability problem (\ref{5n}) and (\ref{6n}), but also 
the original map (\ref{2}) and (\ref{3}), contains convolutions, 
the use of generating functions \cite{Fel}, which 
allows transformations of sums of products into products of sums, could be 
utilized in the investigation of the FSM and some other maps with memory.  
As an example, in the particular case of the stability problem (\ref{5n}) 
and (\ref{6n}), the introduction of the generating functions
\be \label{17gf1}
\tilde{W}_{\alpha}(t)= \frac{K}{\Gamma (\alpha)}
\sum_{i=0}^{\infty }[(i+1)^{\alpha-1}-i^{\alpha-1}]t^i,
\ee
\be \label{17gf2}
\tilde{X}(t)=\sum_{i=0}^{\infty }\delta x_i t^i,
\ee
\be \label{17gf3}
\tilde{P}(t)=\sum_{i=0}^{\infty }\delta p_i t^i,
\ee
leads to
\be \label{17gf5}
\tilde{X}(t)=\frac{p_0 \tilde{W}_{\alpha}(t)}{K}  
\frac{t}{1 - t \Bigl(1- \tilde{W}_{\alpha}(t) \Bigr)  },
\ee
\be \label{17gf4}
\tilde{P}(t)=p_0 \frac{1+  \tilde{W}_{\alpha}(t) }
{ 1-  t \Bigl( 1- \tilde{W}_{\alpha}(t) \Bigr) } .
\ee
Now the original problem is reduced to the problem of the
asymptotic behavior at $t=0$ of the derivatives of the analytic functions
$\tilde{X}(t)$ and $\tilde{P}(t)$, which is still quite complex and is not
considered in this article.

In the region of the parameter space 
where the fixed point is stable, the fixed point is surrounded
by a finite basin of attraction, whose width $W$ depends on the values of $K$
and $\alpha$. For example, for $K=3$ and   $\alpha=1.9$ the width of the 
basin of attraction is $1.6<W<1.7$. 
Simulations of thousands of trajectories with $p_0 < 1.6$ performed by the authors, 
of which only 200 (with $1.5< p_0 < 1.6$) are presented in Fig. 1e, 
show only converging trajectories,
%%Thousands 
%%of simulations with $p_0<1.6$ show only converging trajectories in Fig~1e,
whereas among 200 trajectories with  $1.6<p_0<1.7$ in Fig~2a 
there are trajectories 
converging to the fixed point as well as some trajectories converging to 
attracting slow diverging trajectories (ASDT), whose properties will be
discussed in the following section. Trajectories in Fig.~1e converge very
rapidly. In the case $K=2$ and  $\alpha=1.4$ in addition to the trajectories 
which converge rapidly and ASDTs
there exist attracting slow converging trajectories (ASCT) 
(Fig.~1f).

\section{Attracting slow diverging trajectories (ASDT)}\label{p1} 

As it can be seen from Fig~2a, the phase portrait on a cylinder 
of the FSM with   $K=3$ and   $\alpha=1.9$ contains only one fixed point
and ASDTs approximately equally spaced along the $p$-axis. This result
corresponds to the fact that the standard map with  $K=3$ has only 
one central island. More complex structure of the standard map's phase space 
for smaller values of  $K$  (for example for $K=2$ and $K=0.6$) can explain
more complex structure of the FSM's phase space, where
periodic attracting trajectories with periods $T=4$ (Fig.~2b), $T=2$, 
and $T=3$ (Fig.~2c) are present.

Each ASDT has its own basin of attraction (see Fig.~2d).
Between those basins two initially close trajectories at first diverge,
but then converge to the same or different fixed point or ASDT.

\begin{figure}
\centering
\rotatebox{0}{\includegraphics[width=7 cm]{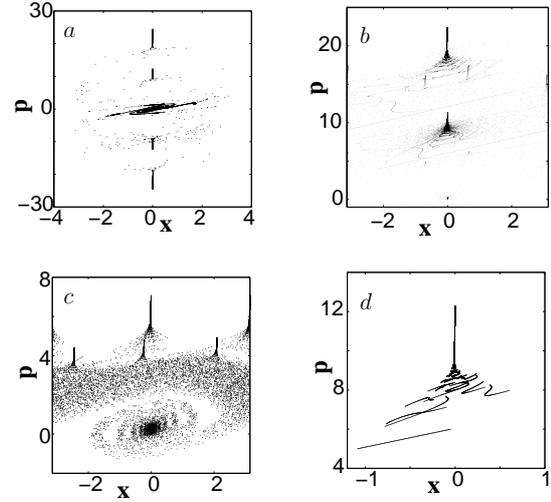}}
\caption{\label{Fig2}  Phase space with  ASDTs: a). The same values of
parameters  as in
Fig~1e  but  $p_0=1.6+0.0005i$; b). 200 iterations on
trajectories with $p_0=4+0.02i$, $0 \le i < 500$ for the case $K=2$,
$\alpha=1.9$. Trajectories converging to the fixed point, ASDTs with $x=0$, and
period 4 attracting trajectories are present;
c). 2000 iterations on
trajectories with $p_0=2+0.04i$, $0 \le i < 50$ for the case $K=0.6$,
$\alpha=1.9$. Trajectories converging to the fixed point, period 2 and 3 
attracting trajectories are present;
d).The same values of
parameters as in
Fig~1e  but  $p_0=5+0.005i$.  
 }
\end{figure}

Numerical evaluation shows that for ASDTs which converge to 
trajectories along the $p$-axis ($x \rightarrow x_{lim}=0$) 
in the area of stability
(which is the same as for the stability of the fixed point) 
the following holds (for large $n$ see Fig.~3a)
\be \label{5i}
p_n = C n^{2-\alpha }.
\ee
The constant C can be easily evaluated for $1.8< \alpha <2$. 
Consider an ASDT with $ x_{lim}=0$, $T=1$, and $2 \pi M$,
where $M$ is an integer,
constant step in $x$ in the unbounded space. 
Then Eq. (\ref{3}) with $b=0$ gives
\be \label{5}
x_{n+1}-x_{n} = \frac{1}{\Gamma (\alpha )} 
\sum_{k=1}^{n} (p_{k+1}-p_k)V_{\alpha}(n-k+1)
%\eqno (5)
\ee
$$
+ \frac{p_1}{\Gamma (\alpha)} V_{\alpha}(n+1) ,
$$
For large $n$ the last term is small ($\sim n^{\alpha-2}$) and the 
following holds
\be \label{6} 
\sum_{k=1}^{n} (p_{k+1}-p_k)V_{\alpha}(n-k+1) = 2 \pi M \Gamma (\alpha).
%\eqno (6)
\ee

\begin{figure}
\centering
\rotatebox{0}{\includegraphics[width=7 cm]{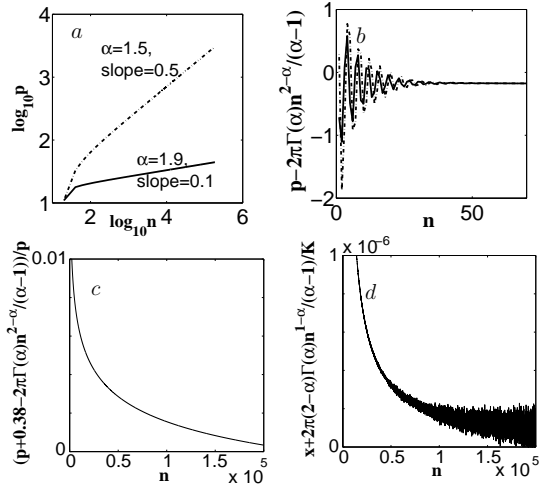}}
\caption{\label{Fig3}  Evaluation of the behavior of the ASDTs: 
a). Momenta for two ASDTs with $x_n \approx 2\pi n$
in the unbounded space (in this example $K=2$). 
The solid line is related to a trajectory with 
$\alpha = 1.9$ and its slope is 0.1. The dashed line  corresponds to a 
trajectory with $\alpha = 1.5$ and its slope is 0.5;
b). Deviation of momenta from the asymptotic formula 
for two ASDTs with $x_n \approx 2\pi n$
in the unbounded space, $\alpha = 1.9$, and  $K=2$. The dashed line 
has $p_0=7$ and the solid one $p_0=6$; 
c). Relative deviation of the momenta for
the trajectories in Fig~3b from the asymptotic formula;
d).  Deviation of the $x$-coordinates  for
the trajectories in Fig~3b from the asymptotic formula.
 }
\end{figure}

With the assumption $p_n \sim  n^{2-\alpha}$ it can be shown that 
for values of $\alpha>1.8$ considered the terms in the last
sum with large $k$ are small and in the series representation
of $V_{\alpha}(n-k+1)$ it is possible to keep only terms of the 
highest order in $k/n$. 
Thus, (\ref{6}) leads to the approximations 
\be \label{7} 
p_n \approx p_0 +  \frac{2 \pi M \Gamma (\alpha) n^{2-\alpha}}{\alpha-1},
\ee
\be \label{7x} 
x_n \approx -\frac{2\pi M(2-\alpha ) \Gamma (\alpha) }{ K(\alpha-1) n^{\alpha-1}}.
\ee
In the case $K=2$, $\alpha=1.9$ Figs.~3b-3d show two  trajectories with $M=1$
(initial momenta $p_0=6$ and  $p_0=7$) approaching an ASDT: the deviation
from the asymptotic (\ref{7}) and (\ref{7x})  
and the relative difference with respect to (\ref{7}).

\section{Period 2 stable trajectory}\label{p2}

The SM has two stable points of the period $T=2$ trajectory for
$4<K<2\pi$ with the property 
\be \label{8} 
p_{n+1} = -p_n, \    \  x_{n+1} = -x_n.
\ee
The same points persist in the numerical experiments for the FSM
(Fig~4a). These
points are  attracting most of the trajectories with small $p_0$. 
Assuming the existence of a $T=2$ attracting trajectory, 
it is possible to calculate the coordinates of its attracting 
points $(x_l, p_l)$ 
and $(-x_l, -p_l)$.
In this case from (\ref{2}) and (\ref{3})

\be \label{9} 
p_l = \frac{K}{2} \sin(x_l),
\ee

\be \label{10} 
x_{l} = \frac{K}{2 \Gamma(\alpha)} \sin(x_l) \sum_{k=1}^{\infty} (-1)^{k+1}
V_{\alpha}(k)
\ee
\begin{figure}
\centering
\rotatebox{0}{\includegraphics[width=7 cm]{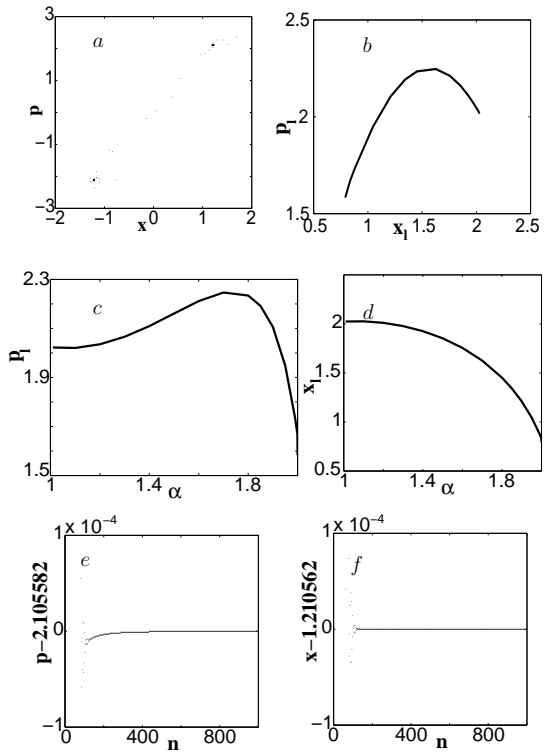}}
\caption{\label{Fig4}  Period 2 stable trajectory: 
a). An example of $T=2$ attractor for $K=4.5$, $\alpha =1.9$.
One trajectory with $x_0=0$,  $p_0=0.513$;
b). $p_l$ of $x_l$ for the case of $K=4.5$; 
c).  $p_l$ of $\alpha $ for the case of $K=4.5$;
d). $x_l$ of $\alpha $ for the case of $K=4.5$;
e). $p_n-p_l$ for the trajectory in Fig.~4a. After 1000 iterations 
$|p_n-p_l| < 10^{-7}$;
f). $x_n-x_l$ for the trajectory in Fig.~4a. After 1000 iterations 
$|x_n-x_l| < 10^{-7}$.
 }
\end{figure}
Finally, the equation for $x_l$ takes the form
\be \label{11} 
x_l = \frac{K}{2 \Gamma(\alpha)} V_{\alpha l} \sin(x_l),
\ee
where 
\be \label{12} 
 V_{\alpha l}  =  \sum_{k=1}^{\infty} (-1)^{k+1} V_{\alpha}(k)
\ee
and can be easily calculated numerically. 
%(see Appendix).
From (\ref{11}) the condition of the existence of $T=2$ trajectory  
\be \label{13} 
K > K_c(\alpha) = \frac{2 \Gamma(\alpha)}{V_{\alpha l}},
%\eqno (13)
\ee
is exactly opposite to  (\ref{17n}). It is satisfied above the curve
$K=K_c(\alpha )$ on the Fig.~1a.
For $\alpha=2$  (\ref{13}) 
produces the well-known condition $K>4$ for the SM. 
The results of calculations of the $x_l$ and $p_l$ 
for the cases $K=4.5$, $1<\alpha<2$ presented in Fig.~4b-d
perfectly coincide with the results of the direct 
computations of (\ref{2}) and (\ref{3}) with $b=0$. 
After 1000 iterations presented in Figs.~4e,f  the values of deviations  
$|p_n-p_l|$ and  $|x_n-x_l|$ are less than $10^{-7}$.

\section{Cascade of bifurcations type trajectories (CBTT)}

Period 2 stable trajectories have limited basins of attraction. Trajectories that don't
fall into those areas reveal a diverse variety of properties,
from period two slow attracting trajectories to fractal type attractors
and cascade of bifurcations type trajectories (CBTT). 
\begin{figure}
%\centering
\rotatebox{0}{\includegraphics[width=3 cm]{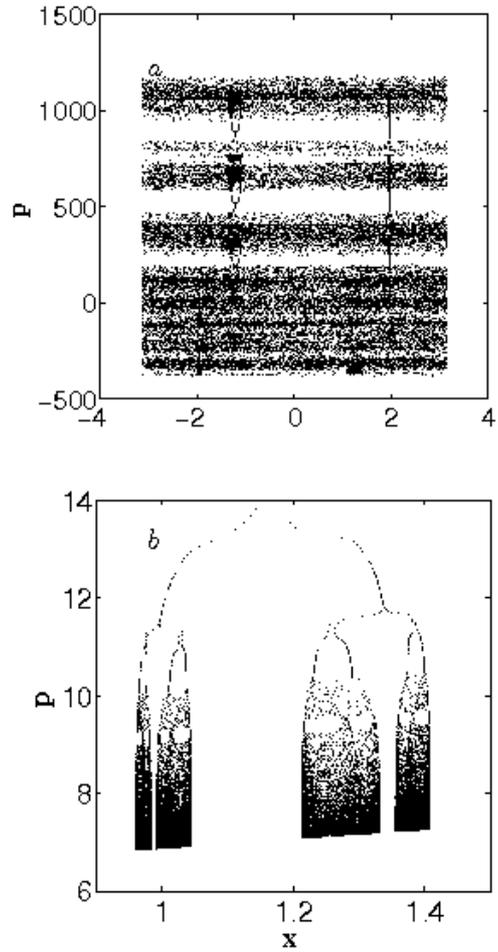}}
\caption{\label{Fig5}  Cascade of bifurcation type trajectories: 
a). 120000 iterations on a single trajectory with $K=4.5$, 
$\alpha=1.65$, $p_0=0.3$. The trajectory occasionally sticks to 
a CBTT but then always recovers into the chaotic sea; 
b). 100000 iterations on a trajectory with $K=3.5$, 
$\alpha=1.1$, $p_0=20$. The trajectory very fast turns into a CBTT
which slowly converges to a fractal type area.
 }
\end{figure}
Fig.~5a presents a single chaotic trajectory which sticks to the areas similar to
the cascade of bifurcations
which are well-known for the logistic map.  
In Fig.~5b a single trajectory falls very rapidly into
one of the attracting CBTTs. Because the bifurcation diagram 
of the logistic map has fractal properties 
(see for example Chapter 2 in \cite{GL}), it is expected that
the structure to which this trajectory slowly converges also possesses
fractal features.

The properties of this type of attractors, as well as 
the properties different types of observed during computer simulations 
chaotic, attracting, and ballistic trajectories
for $K>2\pi $ (see Fig~6) will be considered in the subsequent article.

\begin{figure}
\centering
\rotatebox{0}{\includegraphics[width=7 cm]{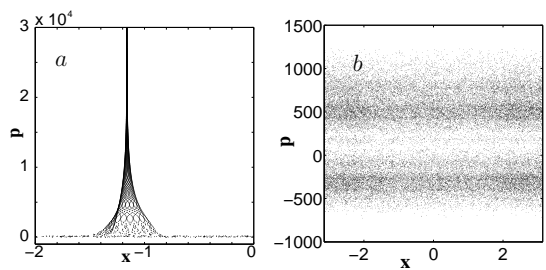}}
\caption{\label{Fig6} Examples of  phase space for $K>2\pi$: 
a). An attracting ballistic trajectory with $K=6.908745$, $\alpha=1.999$,
$p_0=0.7$;
b). A chaotic trajectory for $K=6.908745$, $\alpha=1.9$.}
\end{figure}

\section{Fractional attractors and their stability}

The problems of existence and stability of the fractional attractors 
for the systems described by the FDEs were addressed in a few recent
papers.  It was noticed in \cite{ZSE} that the properties of the fractional 
chaotic attractors are different from the properties of the
``regular'' chaotic attractors and may have some pseudochaotic features. 
The problem of existence of multi-scroll fractional 
chaotic attractors was considered in \cite{TH}. The problem of stability of 
the stationary solutions (fixed points for ODEs) of systems described by the
fractional ODEs and PDEs was considered in \cite{Guf1,Guf2,Guf3}. In the
above mentioned articles the equations contained the Caputo fractional
derivatives, whereas in the present article the Riemann-Liouville 
fractional derivative is used. This fact does not allow a direct 
comparison of the results. The results \cite{ZSE,TH,Guf1,Guf2,Guf3} 
were supported by a relatively small number of computations and this is 
understandable, taking into account all the difficulties of performing 
numerical simulations for the equations with fractional derivatives.

The use of the FSM, which is equivalent to the original FDE, 
allows performing thousands of runs of simulations of the kicked
fractional system with two parameters: $K$ and $\alpha$. The FSM also allows 
making some analytic deductions and revealing some properties of the 
fractional attractors which were not reported before:

a). The stability of the fixed point (0,0) of the FSM is different 
not only from the stability of the fixed point in the domain of the 
regular motion (zero Lyapunov exponent) of the SM, but also from the
stability of fixed attracting points of the regular (not fractional) 
dissipative systems like, for example, the dissipative standard map 
(Zaslavsky map) \cite{ZasMap}.  
The difference is in the way in which trajectories approach the attracting 
point. In the FSM this way depends on the initial conditions. 
For example, in Fig.~1f there are two trajectories approaching the same 
fixed point: one is fast spiraling into the attractor and 
the other is slowly converging. 

b). Stable period 2 attracting trajectories exist only in the asymptotic 
sense - they do not represent any real periodic solutions. 
If the initial condition is chosen in a period two stable attracting
point, this trajectory will immediately jump out of this point and where 
it will end depends on the values of $K$ and $\alpha$.

c). All the FSM attractors exist in the sense that there are trajectories 
which converge into those attractors. But if an initial condition is taken
on any of the attracting trajectories (except for the fixed point), they will 
most likely not evolve along the same trajectory.

\section{Conclusion}

In this article properties of the phase space of the FSM were investigated. It was
shown that islands of regular motion of the SM in the FSM turn into attractors
(points, attracting trajectories, and
fractal-like structures). Properties of the attracting fixed points, period two
trajectories, ASCTs, and ASDTs were considered.
This consideration allows the description of the evolution of the dynamical variable
$x$ of the original fractional dynamical system, a system described by the FDE
reducible to the FSM.  
Physical interpretation of the momentum, defined through a fractional derivative
from the variable $x$, is unclear.

The explanation of the CBTTs, which are interesting phenomena, requires 
further  detailed
investigation. Chaotic trajectories that spend some time near CBTTs, 
which can be
called "sticky attractors" in analogy to  "sticky islands" 
of the SM,  are good
candidates for the investigation of anomalous diffusion. Transport was not
considered in this article.
How general the properties of the phase space of the FSM are will become clear after
further investigations of different fractional maps, maps with memory which can be
derived from the FDEs, and particular those suggested in \cite{Main}, will be
conducted.
The fact that so many physical systems can be reduced to studying of the SM gives a
hope that those physical systems which can be reduced to studying the FSM will be
found.

\section*{Acknowledgments}
We express our gratitude to H. Weitzner 
for many comments and helpful discussions.
The authors thank A. Kheyfits for suggesting the use of generating functions to
solve the FSM fixed point stability problem.
This work was supported by the Office of Naval Research,
Grant No. N00014-08-1-0121.

%\section*{Appendix}

%The function (\ref{12}) can be written as

%\be \label{1A} 
%V_{\alpha l} =  \sum_{k=1}^{\infty} (-1)^{k+1} V_{\alpha}(k) = S_1+S_2,
%\ee
%where
%\be \label{2A} 
%S_1 =  \sum_{k=1}^{2N} (-1)^{k+1} V_{\alpha}(k) 
%\ee
%with the $N$ sufficiently large and 
%\be \label{3A} 
%S_2 =  \sum_{k=N+1}^{\infty} \{V_{\alpha}(2k-1)-V_{\alpha}(2k)\} \   \ .
%\ee

%The value of $S_1$ can be calculated numerically with high precision directly. 
%The second sum can be developed into a series used for computation
%\be \label{4A}
%\begin{gathered}
%S_2 = \sum_{k=N+1}^{\infty}
%(2k)^{\alpha-3}(\alpha-1)(2-\alpha)\Bigl(1+\frac{3-\alpha}{2} \frac{1}{k}+
%\frac{7(3-\alpha)(4-\alpha)}{48}\frac{1}{k^2}+  \\
%\frac{(3-\alpha)(4-\alpha)(5-\alpha)}{32}\frac{1}{k^3}+O(\frac{1}{k^4})\Bigr)
%= \\
%(2)^{\alpha-3}(\alpha-1)(2-\alpha)\Bigl(\zeta(3-\alpha)+\frac{3-%\alpha}{2}\zeta(4-\alpha)+
%\frac{7(3-\alpha)(4-\alpha)}{48}\zeta(5-\alpha)+  \\
%\frac{(3-\alpha)(4-\alpha)(5-\alpha)}{32}\zeta(6-\alpha)+O(\frac{1}{k^{6-%\alpha}})\Bigr)-
%\\
%\sum_{k=1}^{N}
%(2k)^{\alpha-3}(\alpha-1)(2-\alpha)\Bigl(1+\frac{3-\alpha}{2} \frac{1}{k}+
%\frac{7(3-\alpha)(4-\alpha)}{48}\frac{1}{k^2}+  \\
%\frac{(3-\alpha)(4-\alpha)(5-\alpha)}{32}\frac{1}{k^3}\Bigr).
%\end{gathered}
%\ee

%%%%%%%%%%%%%%%%%%%%%%%%%%%%%%%%%%%%%%%%%%%%%%%%%%%%%%%%%%%%%%%%%%%%%%%%%%%%%%%%%

\end{document}